\def \SAIT #1 #2 {{\em Mem.\ Soc.\ Astron.\ It.\/} {\bf #1}, #2}
\def \MESS #1 #2 {{\em The Messenger\/} {\bf #1}, #2}
\def \ASTRNACH #1 #2 {{\em Astron. Nach.\/} {\bf #1}, #2}
\def \AAP #1 #2 {{\em Astron. Astrophys.\/} {\bf #1}, #2}
\def \AAL #1 #2 {{\em Astron. Astrophys. Lett.\/} {\bf #1}, L#2}
\def \AAR #1 #2 {{\em Astron. Astrophys. Rev.\/} {\bf #1}, #2}
\def \AAS #1 #2 {{\em Astron. Astrophys. Suppl. Ser.\/} {\bf #1}, #2}
\def \AJ #1 #2 {{\em Astron. J.\/} {\bf #1}, #2}
\def \ANNREV #1 #2 {{\em Ann. Rev. Astron. Astrophys.\/} {\bf #1}, #2}
\def \APJ #1 #2 {{\em Astrophys. J.\/} {\bf #1}, #2}
\def \APJL #1 #2 {{\em Astrophys. J. Lett.\/} {\bf #1}, L#2}
\def \APJS #1 #2 {{\em Astrophys. J. Suppl.\/} {\bf #1}, #2}
\def \APSS #1 #2 {{\em Astrophys. Space Sci.\/} {\bf #1}, #2}
\def \ASR #1 #2 {{\em Adv. Space Res.\/} {\bf #1}, #2}
\def \BAIC #1 #2 {{\em Bull. Astron. Inst. Czechosl.\/} {\bf #1}, #2}
\def \JSQRT #1 #2 {{\em J. Quant. Spectrosc. Radiat. Transfer\/} {\bf #1}, #2}
\def \MN #1 #2 {{\em Mon. Not. R. Astr. Soc.\/} {\bf #1}, #2}
\def \MEM #1 #2 {{\em Mem. R. Astr. Soc.\/} {\bf #1}, #2}
\def \PLR #1 #2 {{\em Phys. Lett. Rev.\/} {\bf #1}, #2}
\def \PASJ #1 #2 {{\em Publ. Astron. Soc. Japan\/} {\bf #1}, #2}
\def \PASP #1 #2 {{\em Publ. Astr. Soc. Pacific\/} {\bf #1}, #2}
\def \NAT #1 #2 {{\em Nature\/} {\bf #1}, #2}
\def\bge{\begin{equation}}
\def\ede{\end{equation}}

\documentstyle[twoside]{memsait}
\input epsf.sty
\begin{opening}
\title{GALACTIC NUCLEI ACTIVITY SUSTAINED BY GLOBULAR CLUSTER MASS ACCRETION}
\author{ROBERTO CAPUZZO--DOLCETTA$^1$, Paolo Miocchi$^2$}
\institute{$^1$Institute of Astronomy, $^2$Dept. of Physics,
University ``La Sapienza'', Roma, Italy}
\date{} 
\end{opening}

\begin{document}

\oddpagefooter{}{}{} 
\evenpagefooter{}{}{} 
\bigskip

\begin{abstract}
 The decay of globular clusters to the center of their mother galaxy
corresponds to carrying a quantity of mass sufficient to sustain the
gravitational activity of a small pre--existing nucleus and to accrete it
in a significant way. This is due to both {\em dynamical friction} of field stars and {\em tidal disruption} by the compact
nucleus. The results of the simplified model presented here show 
that the active galactic nuclei luminosity and lifetime depend on the
characteristics of the globular cluster system and are quite 
insensitive to the nucleus' initial mass.
\end{abstract}

\section{Introduction}
It is commonly accepted  that the active galactic nuclei (AGN) emission is due to the extraction of energy from the gravitational potential
in form of mass falling onto a super--compact object.
Most of the theoretical study is devoted to the geometry of the problem 
and to physical details in the attempt to explain some spectral 
characteristics and peculiarities of particular objects, rather than
to answer the fundamental question:
\par  --what is  the actual source of the accreting mass? 
\par  Here we try to give 
an answer to this question, carrying evidence 
that spherical mass accretion on a compact object in form of dynamically
decayed globular clusters in a (triaxial) elliptical galaxy can be the source of
the power released by AGNs without invoking ad hoc assumptions (see also Capuzzo--Dolcetta, 1993; Capuzzo--Dolcetta, 1997).
\section{Observed spatial distributions of globular clusters}
It is ascertained that
the radial distribution of globular clusters in their parent galaxy is less peaked than that of bulge stars.  For instance, recent Hubble Space Telescope observations of 14 elliptical galaxies (Forbes et al. 1996) confirm the general trend of flattening of the cluster distribution within $~ 2.5$ Kpc from the centre. 
This difference may be due to different initial conditions or to the 
evolution of the cluster system  distribution (the bulge does not evolve
due to its very long two--body relaxation time).  This latter hypothesis
is  clearly the most ``economical''  and it has been already positively checked
by Capuzzo--Dolcetta (1993).  Here we mainly point the attention to one of
the consequences that the evolution of the globular cluster system
(GCS) has in terms of mass carried to the galactic centre.

\section{Evolution of the globular cluster system}
While the details of the general physical model can
be found in Pesce, Capuzzo--Dolcetta and Vietri 1992,   Capuzzo--Dolcetta 1993
 and  Capuzzo--Dolcetta  (1997),
here we  just recall that the GCS evolution
in an elliptical  galaxy is due to the actions of {\it dynamical friction}
($df$) by field stars and of the {\it tidal} ($tid$) interaction with a compact object in the centre of the galaxy. 

Dynamical friction is the decelerating effect due to the fluctuating
gravitational field of the about $10^{11}$ stars of the galaxy where
the massive globular cluster is orbiting. It is well known  (see, e.g.,
Chandrasekhar 1943) that the  cumulative  gravitational encounters of the satellite with much lighter stars  result in a loss of both energy and
angular momentum of the massive satellite.  Detailed numerical integrations of
globular cluster orbits in a triaxial potential for a set of initial conditions 
(see Capuzzo--Dolcetta 1993) allowed us to describe accurately the effects of dynamical friction on the orbital evolution of a GCS composed by clusters of
different masses.
Moreover, the time rate of energy
and angular momentum decay is proportional to the satellite mass.

The destruction of clusters due to strong tidal interaction with a massive nucleus is,
conversely, more effective for looser (so, usually, lighter) clusters (see Ostriker, Binney and Saha 1988).

As a consequence, the dynamical friction decay and tidal destruction time scales ($\tau_{df}$, $\tau_{tid}$) depend differently on
the individual cluster mass and mass density. Of course,
they also depend 
on the initial cluster energy, as well as  on  galactic potential and velocity distribution together with the nucleus mass ($M_n$).
There is  a competition between the two processes: dynamical friction is
 effective until a nucleus is accreted in mass enough
to destroy the incoming  clusters. This is shown by the 
expression of the ratio of the two time scales ($\overline{\rho}_h$ is 
the cluster density averaged over its half--mass radius)

\begin{equation}
{\tau_{tid}\over \tau_{df}}=f(E)\sqrt{\overline{\rho}_h}{M\over M_n},
\end{equation}

where  $\overline{\rho}_h$ is 
the cluster density averaged over its half--mass radius,
$M$ is the mass of the cluster, $M_n$ is the nucleus mass and $f(E)$ is
a known function of the cluster orbital energy.

The possibility of a self--regulated nucleus formation in a galaxy 
is, so, quite natural. 
Of course several questions have to be answered: 
\par (i) when  are $\tau_{tid}$ and $\tau_{df}$  sufficiently short to be
relevant in the life of a galaxy?
\par (ii) is it possible to build up a compact nucleus in the galactic centre
in the form of dynamically decayed globulars?
\par (iii) what fraction of the mass of frictionally decayed and 
tidally destroyed clusters is swallowed in the compact nucleus?
\par (iv) can a quasar--like emission be explained by such kind
of spherical accretion?
\par A definite quantitative answer to question (ii) will require an 
accurate modelization. Such a modelization is in preparation, 
and it will consist of N-body simulations of the evolution of a set of globular clusters with different initial mass and internal mass spectrum along their orbits
in a galactic triaxial field in presence of a central compact nucleus
(Capuzzo--Dolcetta and Miocchi 1997). This will allow to see under which conditions a globular cluster can release stars to be swallowed by the nucleus,
so to feed its activity.
The questions  (i), (iii) 
(as well as  (iv)) have been answered  
in Pesce et al.  (1992), Capuzzo--Dolcetta  (1993), Capuzzo--Dolcetta  (1997),
and will not be discussed  here. 
\par In what follows we limit ourselves to give support to a positive answer to question (iv).

\section{The model}
A black hole of given mass $m_{bh}$ which stays at the galactic centre
is able to swallow the surrounding stars entering a destruction radius
$r_d=max (r_S,r_t)$ where $r_S$ is the Schwarzschild's radius and
$r_t$ the tidal--breakup radius, defined as

$$
r_t=\left({ {m_{bh}\over <M_*>} }\right)^{1\over 3} <R_*>,
$$

with $ <M_*> $  and  $<R_*>$  mass and radius of the typical star.
 The resulting rate of (spherical) mass accretion is $\dot m_{bh}$, 
which yields a gravitational luminosity 

$$
L_n=\dot m_{bh} \phi
$$

where $\phi$ is the gravitational potential near $r_d$.
The mass accretion rate and the corresponding luminosity  crucially dependson the star density 
($\rho_*$) and velocity dispersion ($<v_*^2>^{1/2}$) around the black hole
through 

\bge
\dot m_* =-\sigma_*\rho_*<v_*^2>^{1/2} \\
\ede
\bge
\sigma_* = \pi r_d^2\left({  1+{ {Gm_{bh}\over r_d} 
\over { {1\over 2} <v_{*}^2> }} }\right) \\
\ede
\bge
\dot m_{bh} =-\dot m_* \\
\ede
\bge
L_n =\eta \dot m_{bh}c^2
\ede

where $\dot m_*$ is the rate of stellar mass swallowed in the black hole
and $\eta$ is an efficiency factor, of the order of $10\%$.
Of course  higher $\rho_*$ corresponds to higher $\dot m_*$,
while a high stellar velocity dispersion $<v_*^2>^{1/2}$ favours nucleus accretion increasing the
capture time rate but,  at the same time, decreases the swallowing
cross section (3).
\par  Capuzzo--Dolcetta (1997) showed that  the 
nucleus accretion rate can easily increase up to 
few $M_\odot yr^{-1}$ due to stars of
 frictionally decayed globular clusters and 
to tidally disrupted clusters.  This accretion rate is exactly of the order 
of that needed to sustain a typical quasar activity (note that the high velocity
bulge stars cannot provide  for more than 
$10^{-7}-10^{-6} M_\odot yr^{-1}$).

\section{Results}
Here we report of some results of a model where the GCS
is composed by clusters all of the same mass and having an initial
density distribution and velocity dispersion equal to those of  bulge stars'  in  a typical triaxial galaxy. 
\par
Figure 1 a) shows the time evolution of the nucleus luminosity 
for different choices of the initial black hole mass  $m_{bh0}$
($10$,  $10^2$,  $10^3$,  $10^6$ $M_\odot$). It is 
 remarkable how,
independently of  $m_{bh0}$ ,  the highest luminosity
reached, $L_{max}$, is similar ($L_{max}$ varies for a factor 
of $10$ when $m_{bh0}$ ranges between $10$ to $10^6$ $M_\odot$).
The nucleus luminosity has
a short super--Eddington burst, a factor $10^3$ brighter than the 
luminosity when a 
relatively slow  dimming phase starts (the noise in the figure is real, due to 
the graininess of the problem amplified by the non--linear coupling of 
source and sink term in the dynamical equations).  Of course  $L_{max}$  is 
attained later for smaller  values of $m_{bh0}$, for
it requires longer to accrete enough mass  onto the nucleus. 
In all the cases the nucleus starts brightening   at 
about $1$ Gyr.
At that age enough  massive globulars have frictionally decayed to
the centre and released stars to the black hole. A flatter  slope of $L_n(t)$
follows (at least for $m_{bh0}\geq 10^2$ $M_ \odot$), due to that the less massive incoming clusters feed less the nucleus.
The nucleus continues  increasing its mass  and becomes, rather quickly,
an efficient tidal destroyer of clusters capable  (in some cases) to keep withiin its potential
well a large fraction of the dispersed cluster' stars, which are eventually
swallowed. 
Figure 1 b) shows that the central black hole mass stabilizes around a 
value ($\approx 5\times 10^8$ M$_\odot$ in this model) which is about the 
same irrespectively of its initial value.
\begin{figure}
\epsfysize=8cm 
\hspace{2.5cm}\epsfbox{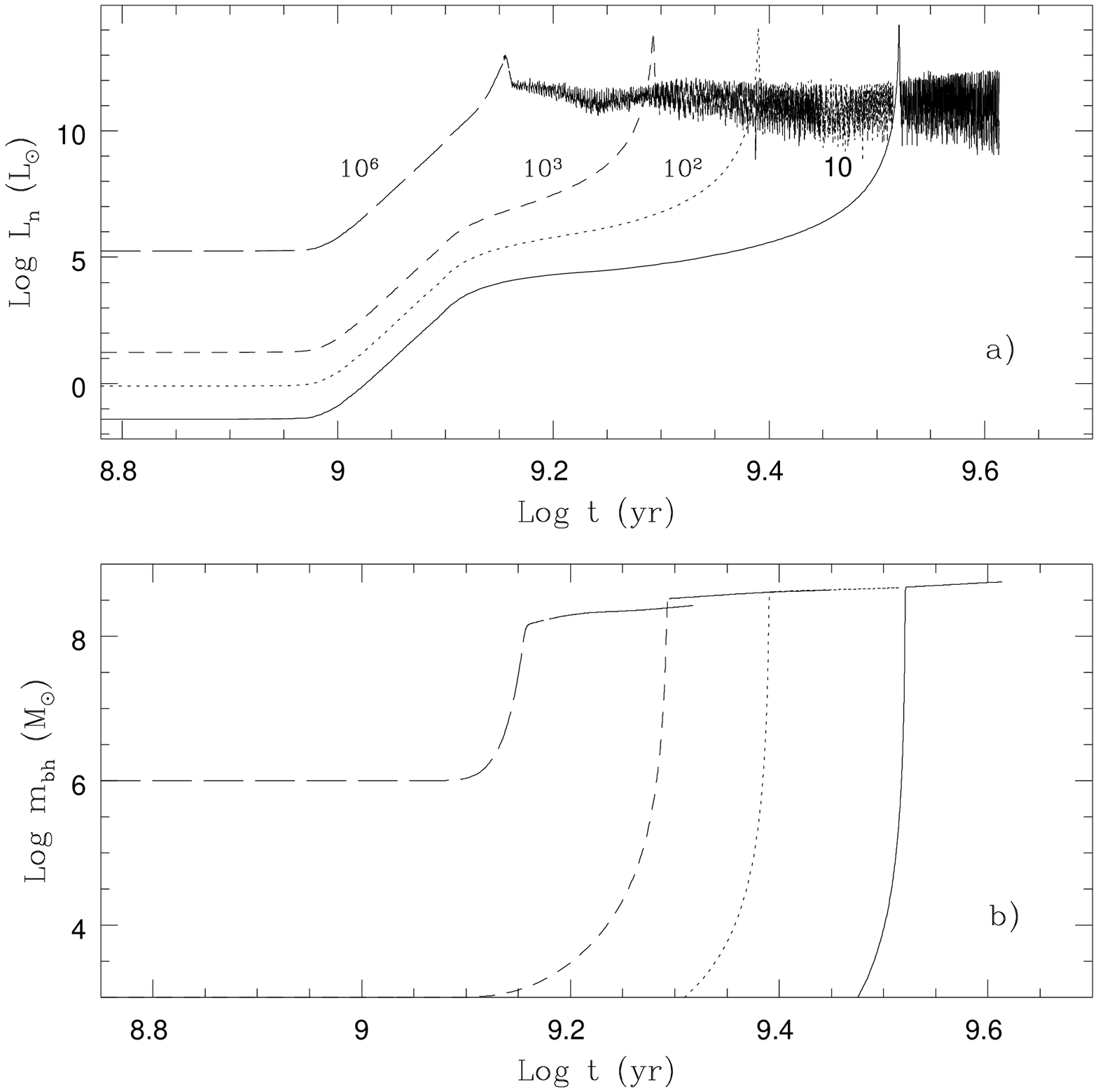} 
\caption{}{panel a): time evolution of  the  nucleus luminosity, for 
various initial nucleus masses (as labeled in $M_\odot$)
  ;  panel b):  time evolution of the nucleus mass, for the same initial
values  as in panel a).}
\end{figure}
 
\section{Conclusions}
A possible mechanism to accrete a compact nucleus 
in the centre of a triaxial galaxy is the swallowing of surrounding stars, which belong
either to dynamically decayed clusters either to tidally destroyed ones. 
 Quantitatively speaking, the relevance of such phenomenon depends on the orbital structure of the GCS
and on its mass and internal density spectrum: a low-velocity dispersion system 
composed by massive, dilute clusters is the most affected by both
dynamical friction and tidal erosion.
In the model here presented,  and
deeply discussed in Capuzzo--Dolcetta (1997), 
the peak of the rate of mass accretion and the final value of the nucleus mass
are remarkably  independent (for a fixed typical globular cluster mass) of the initial black hole mass. What can vary is the age of occurrence of the 
resulting luminosity peak ($10^{13} \div 10^{14}$ L$_\odot$) 
and of the flattening of the mass growth curve.  
Infact, we   found that the black hole mass grows rapidly
until it reaches a value of few $10^8$ M$_\odot$; after that the mass
accretion is much slower.

\end{document}